# MAGNIFICATION BIAS IN GRAVITATIONAL ARC STATISTICS

G. B. Caminha
Centro Brasileiro de Pesquisas Físicas, Rio de Janeiro, Brazil

J. Estrada
Fermi National Accelerator Laboratory, Batavia, Illinois, USA

and

M. Makler
Centro Brasileiro de Pesquisas Físicas, Rio de Janeiro, Brazil
*Draft version September 17, 2013*

## ABSTRACT

The statistics of gravitational arcs in galaxy clusters is a powerful probe of cluster structure and may provide complementary cosmological constraints. Despite recent progresses, discrepancies still remain among modelling and observations of arc abundance, specially regarding the redshift distribution of strong lensing clusters. Besides, fast "semi-analytic" methods still have to incorporate the success obtained with simulations. In this paper we discuss the contribution of the magnification in gravitational arc statistics. Although lensing conserves surface brightness, the magnification increases the signal-to-noise ratio of the arcs, enhancing their detectability. We present an approach to include this and other observational effects in semi-analytic calculations for arc statistics. The cross section for arc formation ($\sigma$) is computed through a semi-analytic method based on the ratio of the eigenvalues of the magnification tensor. Using this approach we obtained the scaling of $\sigma$ with respect to the magnification, and other parameters, allowing for a fast computation of the cross section. We apply this method to evaluate the expected number of arcs per cluster using an elliptical Navarro–Frenk–White matter distribution. We include the effect of magnification using two methods: by considering the magnification dependent arc cross section and by assuming that all sources in the arc forming region are magnified by the same amount. We show that both methods are in excellent agreement and it is thus sufficient to consider the effect of the mean magnification in arc statistics. Our results show that the magnification has a strong effect on the arc abundance, enhancing the fraction of arcs, moving the peak of the arc fraction to higher redshifts, and softening its decrease at high redshifts. We argue that the effect of magnification should be included in arc statistics modelling and that it could help to reconcile arcs statistics predictions with the observational data.

*Subject headings:* gravitational lensing: strong, galaxies: clusters: general, dark matter, cosmological parameters

## 1. INTRODUCTION

Gravitational arcs are highly distorted images of background galaxies (sources) due to the gravitational field of a foreground object (lens), usually a galaxy or cluster of galaxies. After the discovery of the first gravitational arcs (Lynds & Petrosian 1986; Soucail et al. 1987), it was soon realized that the statistics of arcs in galaxy clusters could provide powerful probes of both cluster structure (see, e.g., Wu & Hammer 1993; Bartelmann & Weiss 1994; Hattori, Watanabe, & Yamashita 1997; Oguri, Lee, & Suto 2003) and cosmology (see, e.g., Grossman & Saha 1994; Wu & Mao 1996; Bartelmann et al. 1998; Zieser & Bartelmann 2012). For a complete review on arc statistics, see Meneghetti et al. (2013).

With this in mind, several arc searches were undertaken, both in imaging surveys (Gladders et al. 2003; Cabanač, et al. 2007; Estrada et al. 2007; Faure et al. 2008; Belokurov et al. 2009; Kubo et al. 2010; Gilbank et al. 2011; More et al. 2012; Wen et al. 2011; Bayliss 2012; Wiesner et al. 2012; Pawase et al. 2012; Erben et al. 2013, More et al., in prep, Caminha et al., in prep.), as well as in fields targeting X-ray and optical clusters, with observations from the ground (Luppino et al. 1999; Zaritsky & Gonzalez 2003; Campusano et al. 2006; Hennawi et al. 2008; Kausch et al. 2010; Furlanetto et al. 2013b) and from space (Ebeling, Edge, & Henry 2001; Smith et al. 2005; Sand et al. 2005; Horesh et al. 2010; Okabe et al. 2013). With these surveys, the homogeneous samples of arcs are reaching the order of hundreds of objects.[1] In the near future, the next generation wide-field surveys such as the *Dark Energy Survey*[2] (DES; The Dark Energy Survey Collaboration 2005; Annis, et al. 2005; Ngeow, et al. 2006, Lahav et al., in preparation, Frieman et al., in preparation), which started its first observing season this year, are expected to lead to the discovery of thousands of gravitational arcs.

In parallel to the growing size of arc samples, the theoretical modelling of arc statistics has also experienced significant improvements. The earliest predictions for arc abundance used idealised lenses (from axially symmetric singular isothermal models to halos from low resolution pure CDM simulations) and sources (constant surface brightness sources distributed in a single plane) and strongly underestimated the number of arcs in the now standard ΛCDM cosmology with respect to observational data (see, e.g., Wu & Hammer 1993; Wu & Mao 1996; Bartelmann et al. 1998). Several degrees of realism were added subsequently, investigating the role of each in arc statistics. For example, on the lens side, the effects of cluster ellipticity, asymmetries, and substructures (e.g., Hennawi, Dalal, Bode, & Ostriker 2007; Meneghetti et al. 2007), triaxiality (Oguri, Lee, & Suto 2003; Dalal, Holder, & Hennawi 2004; Meneghetti et al. 2010), and mergers (Torri et al. 2004; Fedeli et al. 2006; Hennawi, Dalal, Bode, & Ostriker 2007; Redlich et al. 2012) were investigated. The effects of baryonic processes of cooling (Puchwein et al. 2005; Rozo, Nagai, Keeton, & Kravtsov 2006; Wambsganss, Ostriker & Bode 2008), star formation rate, and AGN feedback (Mead et al. 2010; Killedar et al. 2012) on the cluster structure, as well as the contribution of cluster galaxies (Flores, Maller & Primack 2000; Meneghetti et al. 2000) to the cluster lensing potential were taken into account. The contribution of structures along the line of sight (Hilbert, White, Hartlap, & Schneider 2007; Puchwein & Hilbert 2009) was also addressed. On the sources side, the impact of their size and ellipticity was investigated (e.g., Oguri 2002), including the effect of arc formation by the merging of multiple images (Rozo, Nagai, Keeton, & Kravtsov 2006). The impact of the redshift distribution of sources was also considered (Wambsganss, Bode, Ostriker 2004; Dalal, Holder, & Hennawi 2004; Li et al. 2005).

The inclusion of the effects above helped to shed light on their contribution on arc statistics. Most of them increase the arc incidence up to factors of a few and their inclusion has reduced the overall discrepancy between observed and predicted arc abundances (e.g., Dalal, Holder, & Hennawi 2004; Horesh et al. 2005; Hennawi, Dalal, Bode, & Ostriker 2007). One of the key factors to reduce this discrepancy was the inclusion of realistic source distributions in the theoretical predictions (Horesh et al. 2005; Wu & Chiueh 2006; Hennawi, Dalal, Bode, & Ostriker 2007). Another key progress was the development of simulations at the image level (Meneghetti et al. 2008; Horesh et al. 2011; Boldrin et al. 2012), which led to the inclusion of new levels of realism, such as sources with surface brightness variations and noise, and allowed for a more direct comparison with the observations, including the use of the same tools for arc detection. The study of selection biases also contributed to a more accurate comparison between simulated and real data (Meneghetti et al. 2010). Overall, these developments have, if not solved, at least mitigated the longstanding arc statistics problem (Meneghetti et al. 2011).

If on one hand the simulations are generally beginning to broadly agree with the observational data, on the other hand several open issues remain, among which the variation of arc abundance with respect to the cluster redshift. For example, Gladders et al. (2003) found an over-abundance of arcs in high-redshift clusters as compared to lower redshift ones in the Red Sequence Cluster Survey (RCS). In particular, they found no arcs in clusters at $z \lesssim 0.6$. However, theoretical investigations did not predict such an enhanced incidence of arcs at high $z$. For example, Hennawi, Dalal, Bode, & Ostriker (2007) found no increase in arc abundance at the redshift implied by RCS. Gonzalez et al. (2012) found arcs in a cluster at $z = 1.75$, which should not be present at their image depths according to their modelling.

Up to the present, most of the focus of theoretical studies in this field — from idealized analytical treatments to fully numerical ray tracing methods in simulated mass distributions — has been on the modelling of the lens potential and on gravitational lensing aspects of arc formation in general. However, observational effects such as finite PSF size and the noise present in real images play a fundamental role in arc detection. An important progress to include these effects was undertaken by Horesh et al. (2005), who added noise (sky and detector backgrounds, Poisson noise, and readout noise), PSF, and light from cluster galaxies to the arc simulations (see also Meneghetti et al. 2005, 2008;

---

[1] For an up-to-date listing of known arc systems see `http://admin.masterlens.org/index.php`.
[2] www.darkenergysurvey.org

Horesh et al. 2011; Boldrin et al. 2012). However, the full image simulation process, including the generation of clusters in N-body/hydrodynamical simulations, the computation of deflection angles, the association of images to sources, and the addition of observational effects; together with the detection of arcs in the simulated images, can be extremely expensive in terms of computational time.

An alternative to this "fully numerical" treatment is to use semi-analytical approaches, which can incorporate, to a certain degree, some of the features of the realistic modelling described above. For example, analytic density profiles can be used instead of N-body simulations, local prescriptions to define the formation of arcs instead of finding images of finite sources, etc. Although this approach is expected to be less accurate, it is useful to investigate the qualitative behavior of the arc abundance with respect to several variables. It can also provide a valuable insight to isolate the key factors that contribute to the modelling of arc abundance, being helpful for interpreting and understanding the outcome of simulations. Furthermore, if calibrated with realistic simulations, the semi-analytic modelling may provide fitting formulae for fast parameter estimation (e.g., to obtain cluster parameters from arc statistics) and for producing forecasts.

In this paper, we use a semi-analytic approach to compute the expected number of arcs per cluster as a function of cluster redshift, mass, and ellipticity. We discuss how to include some observational effects and, in particular, we take into account, for the first time in this approach, the effect of the noise in the images. The signal-to-noise ratio ($S/N$) of the arc images is enhanced by the magnification, effectively changing the limiting magnitude of the image in the region of arc formation, enhancing the efficiency of arc detection. This results in a kind of *magnification bias* and leads us to include this quantity in the arc abundance modelling.

We obtain scaling relations that allow for a fast computation of the cross section for arc formation, taking into account its variation with respect to magnification. We apply this method to an elliptical Navarro–Frenk–White matter distribution in clusters and compute the number of arcs per cluster considering a realistic redshift distribution of sources, which is effectively modified due to the effect of the magnification.

The paper is organized as follows. In the next section we present the semi-analytic modelling for the arc fraction in clusters. In section 3 we discuss the importance of the signal-to-noise ratio for arc detection and how to incorporate the effect of $S/N$ enhancement through the magnification in gravitational arc statistics. In section 4 we review the semi-analytic computation of the arc cross section and the lens model that will be used. In section 5 we obtain the scaling of the cross section with respect to the magnification, which is needed to take the noise into account in the modelling. Finally, in section 6 we apply these results to compute the expected number of arcs per cluster — showing the importance of magnification — and discuss its evolution with respect to the cluster redshift. We summarize our results and present some concluding remarks in section 7. In the Appendix we derive useful expressions for the lensing quantities of elliptical lens models for any parameterisation of the ellipticity.

## 2. SEMI-ANALYTICAL MODELLING FOR ARC ABUNDANCE: INCLUDING OBSERVATIONAL EFFECTS

Our aim is to compute the average number of arcs *per cluster* $f_{arcs}$, for clusters with given properties, such as mass and ellipticity (see, e.g., Cypriano et al. 2001; Hattori, Watanabe, & Yamashita 1997; Oguri, Lee, & Suto 2003; Horesh et al. 2005, for other works using this same statistical measure). For a sample of clusters selected with uniform criteria, for example a given richness and redshift interval, this fraction is easily calculated as $f_{arcs} = N_{arc}/N_{cluster}$, where $N_{cluster}$ is the number of clusters in the sample and $N_{arcs}$ is the total number of arcs found in this sample. While many studies on arc statistics have focused on predictions for the total number of arcs, this quantity is not very suitable to probe properties associated with gravitational lensing (such as lens mass profiles or cosmological distances), since it is extremely sensitive to the cosmological parameters in a way not related to lensing at all. More specifically, the total number of arcs depends on cluster abundance, which is in turn very sensitive to $\sigma_8$ and the growth factor. Some inconsistencies among arc abundance predictions were due to the different choices of the cosmological parameters, rather than the modelling of the lensing aspects of the problem (see, e.g., Li et al. 2005). By focusing on $f_{arcs}$ we may decouple the generation of arcs from cluster abundance, minimizing the explicit dependence on cosmology. Moreover, this statistic is also well suited for targeted surveys, were a sample of clusters is selected to be observed with better imaging to search for arcs. Finally, the fraction of arcs is less dependent on the cluster selection function than the total arc number (of course, as long as the selection is not based on the presence of arcs). A careful selection of the clusters for this statistic could also mitigate the selection biases (Meneghetti et al. 2010).

A direct method to evaluate the expected number of arcs per cluster is to perform simulations where a given mass distribution is used to lens objects from a real astronomical image. The resulting images of these sources are identified as arcs, according to their properties. As mentioned in section 1, the source redshift distribution must be taken into account in computing the arc abundance. For this sake, redshifts can be associated to each source in the image, as in Horesh et al. (2005). A similar approach is to lens sources distributed in multiple planes (Hennawi, Dalal, Bode, & Ostriker 2007). Another possibility, which we refer to as "semi-analytic modelling for arc abundance", consists in computing the cross section for arc formation $\sigma$ for a given cluster (i.e., the effective area in the source plane where a source gives rise to an arc in the lens plane) as a function of source redshift $z_S$ and convolving it with the source surface number density $\Sigma_S$ as a function of $z_S$.

The cross section is generally defined in terms of a given set of parameters of a lensed object. For example, a standard criterium, which we shall apply in this work, is to use the length-to-width ratio ($R := L/W$) of the images. Whenever this ratio is above a given threshold ($R_{th}$) the image is said to form an arc. Other parameters that are relevant for detection include the magnitude and, possibly, the curvature. Naturally, $\sigma$ is a function of the lens properties, such as mass $M$, ellipticity $e$, slope of the density profile $\gamma$, and characteristic radius $r_s$, and also depends on the lens and source redshifts, $z_L$ and $z_S$, through the cosmological distances. It may also depend on source properties, such as ellipticity $e_S$ and angular size $\theta_S$. Therefore, we may write the cross section as $\sigma = \sigma\left(\Pi_L, z_L; \Pi_S, z_S; Q_P\right)$, where $\Pi_L$ is a set of cluster properties ($M$, $r_s$, $\gamma$, $e$, etc.), $\Pi_S$ is a set of source properties ($\theta_S$, $e_S$), and $Q_P$ is a set of arc properties ($L/W$, $S/N$, $\mu$, etc.).[3]

Besides its dependence on $z_S$, the source distribution $\Sigma_S$ is a function of $\theta_S$, $e_S$, and magnitude $m$. In principle, $\Sigma_S\left(z_S, \Pi_S\right)$ can be estimated from deep space-based surveys, such as the COSMOS field[4] (see, e.g., Scoville 2007). If we do not take into account the source sizes and ellipticies, $\Sigma_S$ can be estimated from ground based observations.

A third key ingredient is related to the observational effects, which transform the "proper" arc parameters $Q_P$, that would be measured in noise free images without PSF effects, into their observed ones $Q_O$. For example, the seeing will change the width of the arcs in a way that can be modelled approximately by $W_O = \sqrt{W_P^2 + \sigma_s^2}$, where $\sigma_s$ is the seeing FWHM. These effects are encoded in the "transfer" function $T\left(Q_P | Q_O\right)$.

The final important ingredient in the arc fraction calculation is the detection efficiency, which depends only on the observed arc properties $Q_O$ (i.e., of the objects identified in real images, which are PSF convolved and include all sources of noise) in a given survey. Usually it is assumed that all arcs with observed $(L/W)_O$ and peak surface brightness above some threshold will be detected in a given image. However, in realistic arc finding methods the efficiency is a more complicated function of these quantities. The efficiency $P(Q_O)$ can be determined, for each survey, using image simulations containing fake arcs, as in Estrada et al. (2007) (see also Horesh et al. 2005; Furlanetto et al. 2013a).

Taking into account the above aspects, we may express the arc fraction for clusters with properties $\Pi_L$ as

$$f_{arcs}\left(\Pi_L, z_L\right) = \int \frac{\mathrm{d}\sigma\left(\Pi_L, z_L; \Pi_S, z_S; Q_P\right)}{\mathrm{d}Q_P} \frac{\mathrm{d}\Sigma_S\left(z_S, \Pi_S\right)}{\mathrm{d}z_S \, \mathrm{d}\Pi_S} \, T\left(Q_P | Q_O\right) P\left(Q_O\right) \mathrm{d}z_S \, \mathrm{d}\Pi_S \, \mathrm{d}Q_O \ . \tag{1}$$

In practical applications this expression can be simplified. For example, we shall see in section 4 that a scaling of $\sigma$ with $z_L$, and $z_S$ can be taken into account analytically. Besides, in section 5 we obtain a good approximation for the scaling of the cross section with an arc parameter $Q_P$ (in this case, the magnification $\mu$). Finally, in our modelling of section 6, $\sigma$ will be independent of $\Pi_S$. Therefore, the number of degrees of freedom in equation 1 can be substantially reduced, and some of the remaining ones can be included in an analytic form. Furthermore, in many cases we may explicitly write $Q_P$ in terms of $Q_O$ (as the relation among $W_P$ and $W_Q$ discussed above), instead of using $T\left(Q_P | Q_O\right)$.

The formalism described above allows us to take into account some observational effects together with the detection efficiency, which are usually considered only in full image simulations. A useful aspect of this approach is that it naturally separates the problem into various components, enabling investigations on the influence of each part in the arc statistics predictions.

In this work we shall discuss the effects of the signal-to-noise ratio in the images, which leads us to consider the role of the magnification in the computation of the arc fraction in clusters.

## 3. THE ROLE OF MAGNIFICATION

Since gravitational lensing conserves surface brightness, it is commonly assumed that the magnification is not important for the detection of extended objects such as arcs. This would indeed be true if no noise was present in the images. However, in a real astronomical situation the total flux may have an important role in the detection, due to the signal-to-noise ratio of the arcs. In fact, arcs will not be identified by their peak surface brightness, but rather by the coherence of the signal produced by that object compared to the noise in the image. Therefore, the signal-to-noise ratio $(S/N)$ must play an important role in arc detection.

Here we assume that the dominant sources of noise are Poisson noise from the background and from the object counts. The signal is simply given by the total flux of the object (counts) and the noise is the integrated noise on the object from the Poisson noise of the background and the object. For an object with surface brightness $F$ and occupying the area $A$ in an image whose background has a surface brightness $b$, this ratio is

$$\frac{S}{N} = \frac{FA}{\sqrt{(b+F)A}}. \tag{2}$$

The gravitational lensing effect changes only the image area, with the new area being $A_{lensed} = \mu \times A$, where $\mu$ is the magnification. Taking this into account, the lensed object has

$$(S/N)_{\text{lensed}} = (S/N)_{\text{unlensed}}\sqrt{\mu}. \tag{3}$$

The enhancement of the signal-to-noise of the lensed objects will lead to an increase of the effective depth of the image in the magnified regions, which will be translated into an effective change of the distribution of sources that can be mapped into arcs. Let us assume that the objects in an image are identified and selected in terms of their $S/N$ (e.g., imposing a cut of $5\sigma$ above noise for detection). In this case, the effective flux limit will decrease by a factor of $1/\sqrt{\mu}$ (c.f. Eq. 3) in comparison to a non-lensed region, due to the effect of magnification. In other words, the effective limiting magnitude will be given by

$$m_{\lim}^{\text{eff}} = 2.5\log(\sqrt{\mu}) + m_{\lim}. \tag{4}$$

---

[3] Of course, the cross section will also depend on the cosmological parameters. However, to simplify the notation, we will not put these parameters explicitly, since we shall not address the cosmological dependence of the cross section in this paper.

[4] http://cosmos.astro.caltech.edu/

To include this effect in gravitational arc statistics, we use the formalism of section 2. The arc detection efficiency $P(Q_O)$, which will be a function of the arc $S/N$, is connected to the source distribution in an intricate way through Eq. (1). The cross section will depend on arc and source properties and the transfer function will map the intrinsic to observed arc properties. Here we will make a few simplifying assumptions as a first approach to the problem. First, we assume that the total signal-to-noise and the length-to-width ratios are the dominant factors to determine the detectability of the arcs. We will further assume that the sensitivity to these quantities is uncorrelated and that all arcs with $L/W$ above a given threshold $R_{\rm th}$ and with $S/N$ above a threshold $(S/N)_{\rm th}$ can be detected. In other words, the selection function will be given by

$$P(Q_O) = \theta\left(L/W - R_{\rm th}\right)\theta\left((S/N) - (S/N)_{\rm th}\right),\tag{5}$$

where $\theta(x)$ is the step function. In reality the situation is a bit more complicated since other factors, such as curvature and surface brightness, may impact the detection efficiency and also the sensitivity to $S/N$ and $L/W$ may be correlated. Furthermore, in realistic situations the selection function $P(Q_O)$ will be dependent on the arc finding method. Here we assume that the object pixels are selected as part of the object identification or arc finding methods. This can be achieved, for example, by requiring a low significance for each object pixel with respect to the background. The $S/N$ is then computed taking into account all object pixels and a higher significance is required for the object to be selected as an arc (provided that the $L/W$ restriction is met). As for the hard cut in $L/W$ in Eq. (5), this is implicitly assumed in most arc statistics calculations. In any case, the novel aspect that we shall address in this work is the dependency on $S/N$, which will be shown to have a large impact on the arc abundance predictions.

We shall also neglect for now the explicit dependence of the arc $L/W$ and $S/N$ on the image PSF, such that the transfer function is given simply by

$$T(Q_P|Q_O) = \delta\left((L/W)_O - (L/W)_P\right)\delta\left((S/N)_O - (S/N)_P\right),\tag{6}$$

where $\delta$ is the Dirac delta function, and the quantities with $O$ refer to the observed ones and those with $P$ correspond to what would be measured without the effects of the PSF (but same instrument and other observing conditions).

A third ingredient in the arc abundance calculation is the source distribution. In this paper we will neglect any dependence of the cross section on the properties of finite sources, such as size and ellipticity (see Sect. 4). Therefore, we only need to consider its dependence on redshift and source $S/N$, i.e. $\Sigma_S(z_S, (S/N)_{\rm source})$. The condition (6) implies that we need to consider all arcs with $S/N$ above a given threshold. This will imply that all sources above a certain $S/N$ (as would be seen in the same observing conditions, but no lensing) may give rise to arcs and the connection among these thresholds is given by equation (3). In other words, we need to consider $\Sigma_S(z_S, > (S/N)_{\rm th}/\sqrt{\mu})$. Here we are of course assuming the same observing conditions for the determination of the source density and the arcs, such that we can use Eq. (3). In practice we only need to know how $\Sigma_S(z_S)$ scales in terms of an $S/N$ threshold. Usually the source distribution determined from imaging surveys is expressed in terms of $z_S$ and a limiting magnitude $m_{\rm lim}$ (as long as $m_{\rm lim}$ is well below the detection limit of the images), not a $S/N$ threshold. However, this limiting magnitude is dependent on the threshold. We mimic this effect following Eq. (4). Therefore, after integrating on the arc properties $Q_O$ in Eq. (1) and applying the selection function (6), we only need to consider the surface density of sources per unit redshift, for a given limiting magnitude, $n\left(z_S, m_{\rm lim}^{\rm eff}(\mu, m_{\rm lim})\right) := {\rm d}\Sigma_S\left(z_S, m_{\rm lim}^{\rm eff}(\mu, m_{\rm lim})\right)/{\rm d}z_S$. An explicit form for this function will be used in Sect. 6.

Finally, we need to consider the cross section and its dependence on source and arc properties. As mentioned above, for simplicity we will not consider the effect of finite sources. Besides, the cross section will not depend on $S/N$. The only dependence on the arc parameters will be on $L/W$ and $\mu$. Also, as in the case of the integral over $(S/N)_O$, the integral over $(L/W)_O$ will lead to a threshold in $(L/W)_P$, $R_{\rm th}$. Thus we only need to compute the cross section for $(L/W) > R_{\rm th}$, which we will refer to as $\sigma(\Pi_L, z_L; z_S; \mu, R_{\rm th})$.

Inserting all these ingredients in Eq. (1) and integrating over the arc parameters $\mu$, $(L/W)_O$, and $(S/N)_O$ we finally obtain the expression for the arc fraction including the effect of magnification

$$f_{arcs}(\Pi_L, z_L) = \int \frac{{\rm d}\sigma(\Pi_L, z_L; z_S; \mu, R_{\rm th})}{{\rm d}\mu}\, n\left(z_S, m_{\rm lim}^{\rm eff}(\mu, m_{\rm lim})\right)\,{\rm d}z_S\,{\rm d}\mu\,.\tag{7}$$

In the next section we discuss the computation of the cross section, and in particular the inclusion of its explicit dependence on the magnification, employing a semi-analytic approach.

## 4. SEMI-ANALYTICAL CROSS SECTION

To compute the cross section for the production of arcs, a simple approach is to resort to the local mapping given by the magnification tensor, which determines the distortion of infinitesimal sources. For this sake, we evaluate the two eigenvalues of the magnification tensor to determine the region in the image plane in which their ratio exceeds a certain threshold $R_{\rm th}$. It is in this region that arcs with $L/W$ exceeding $R_{\rm th}$ are expected to be formed.

The magnification eigenvalues are given by

$$\mu_t = \frac{1}{1 - \kappa - \gamma} \quad \text{and} \quad \mu_r = \frac{1}{1 - \kappa + \gamma},\tag{8}$$

where the convergence $\kappa$ and the shear strength, $\gamma = \sqrt{\gamma_1^2 + \gamma_2^2}$, can be obtained from the derivatives of the lens potential $\Psi$:

$$\kappa = \frac{1}{2}\left(\Psi_{11} + \Psi_{22}\right),\ \gamma_1 = \frac{1}{2}\left(\Psi_{11} - \Psi_{22}\right),\ \text{and}\ \gamma_2 = \Psi_{12},\tag{9}$$

where the subindices 1 and 2 refer to derivatives with respect to $\theta_1$ and $\theta_2$ respectively and $\Psi = 2\psi \, D_{LS}/(D_{OS}D_{OL})$, where $\psi$ is the projected gravitational potential of the lens, $D_{OS}$ denotes the angular diameter distance from the observer to the source, $D_{OL}$ that from the observer to the lens, and $D_{LS}$ that from the lens to the source[5] (see, e.g., Schneider, Ehlers, & Falco 1992; Mollerach & Roulet 2002).

Usually, $\mu_t$ corresponds to a deformation approximately in the tangential direction, while $\mu_r$ corresponds to the deformation along the radial direction. Therefore tangential arcs are expected to be formed in regions where $R := |\mu_t/\mu_r| \gg 1$. For infinitesimal circular sources, the length to width ratio ($L/W$) of tangentially distorted images will be approximately given by the local deformation $R$ (Hattori, Watanabe, & Yamashita 1997; Hamana & Futamase 1997). Arcs are usually defined as objects with axial ratio above a given threshold $L/W > R_{th}$ (a typical value considered is $R_{th} = 10$) and, according to the approximation discussed above, they will be produced in a region $\Omega_L$ around the lens for which the local eigenvalue ratio satisfies the condition $R > R_{th}$.

This region in the image plane is mapped through the lens equation into a region in the source plane. The cross section for arc production $\sigma$ (which we will refer to as *deformation cross section*) will be given by the corresponding solid angle in the source plane. Since the solid angles in the image and source planes are related by $d\Omega_S = d\Omega_L/|\mu(\theta)|$ (where $|\mu| = |\mu_t \times \mu_r|$ is the magnification) we hence have (see, e.g., Blandford & Narayan 1986; Fedeli et al. 2006; Dúmet-Montoya et al. 2012)

$$\sigma(\Pi_L; R_{th}) = \int_{\Omega_L(R > R_{th})} d^2\theta \; \frac{1}{|\mu(\theta)|}. \tag{10}$$

Notice that, through this procedure, two points in the lens plane with $R > R_{th}$ that are mapped into the same point in the source plane are accounted for twice in the computation of $\sigma$. This is the appropriate definition for computing $f_{arcs}$ since this region in the source plane will give rise to two arcs. Hence $\sigma$ is an *effective area* in the source plane, already accounting for the multiplicity of the images.

Throughout this work we model the lenses by an elliptical Navarro, Frenk, & White (1996, 1997, hereafter NFW) mass distribution, whose radial (i.e. angle averaged) profile is given by

$$\rho = \frac{\rho_s}{(r/r_s)\,(1 + r/r_s)^2}, \tag{11}$$

where the parameters $r_s$ and $\rho_s$ can be expressed in terms of the mass $M_{200}$ and the concentration parameter $c$ (see Sect. 6).

The lensing properties may be obtained from the projected mass distribution, which in the case of the NFW profile is determined by two parameters (see, e.g., Bartelmann 1996): $r_s$ and

$$\kappa_s := \frac{\rho_s r_s}{\Sigma_{cr}}, \tag{12}$$

where $\Sigma_{cr} = D_{OS}/(4\pi G D_{OL}D_{LS})$.

An elliptical surface density distribution can be obtained from the projected NFW profile $\Sigma_{NFW}(\xi) := \int_{-\infty}^{\infty} \rho(\xi^2 + z^2)dz$ (Bartelmann 1996) by replacing the radial coordinate $\xi$ by

$$\xi^2 = (1 - e)\,x^2 + \frac{y^2}{1 - e} \tag{13}$$

where the parameter $e$ denotes the ellipticity. With this choice, the projected mass contained inside contours of constant $\xi$ is independent of $e$. The relevant quantities for the lensing mapping can be expressed in terms of integrals, which are given in the Appendix.

Since the lens equation can be written in terms of an arbitrary length scale (see, e.g., Schneider, Ehlers, & Falco 1992), we may work with all distances in units of $r_s$ and eliminate this quantity from the analysis, such that the (rescaled) cross section $\tilde{\sigma}$ becomes a function of $\kappa_s$ and $e$ only (for a given $L/W$ threshold). The cross section in steradians is then given by[6] (Dúmet-Montoya et al. 2012)

$$\sigma(\kappa_s, r_s, e) = \tilde{\sigma}(\kappa_s, e) \left(\frac{r_s}{D_{OL}}\right)^2. \tag{14}$$

This already simplifies the calculation of the cross section by eliminating one degree of freedom from the problem (see, e.g., Oguri 2002; Oguri, Lee, & Suto 2003; Fedeli et al. 2006, for similar transformations). We will refer to $\tilde{\sigma}$ as the dimensionless cross section. To recover the cosmological and mass dependence of the parameters, one has to write $\kappa_s$ and $r_s$ in terms of the NFW model parameters $M_{200}$ and $c$, and the cosmological distances (see Sect. 6).

A similar procedure as the one described above for computing the arc cross section is often used for computing the magnification cross section for point-like sources (although, of course, the integration domain is defined by the constraint $\mu > \mu_{th}$, rather than $R > R_{th}$). For finite-size objects such as arcs, the validity of this method has to be checked. Indeed, due to the extended nature of the images, variations of the local eigenvalue ratio throughout the sources may be important. Source ellipticity will break the direct relation between $L/W$ and $R$. Moreover, it is well known that arcs can be formed as the result of the merging of multiple images. The criterium based on the local eigenvalue ratio can only take into account arcs generated through the high distortion of single images, produced when sources are close to any caustic, but not those formed by image merging (Rozo, Nagai, Keeton, & Kravtsov 2006). Some of these finite source effects can be accounted for in the semi-analytic computation of the cross section. For example Fedeli et al. (2006) introduce a method to include the gradients of $R$ across the source. Keeton (2001c) proposed an approximation to obtain $L/W$ from the magnification eigenvalues considering source ellipticity and orientation.

---

[5] Throughout this paper we shall use units in which the light speed is unity.

[6] Throughout this work we will set $R_{th} = 10$, so that this value is implied whenever this variable is not shown explicitly, to alleviate the notation. In the NFW model, the only lens parameters $\Pi_L$ relevant for gravitational lensing are the characteristic convergence $\kappa_s$, scale radius $r_s$, and projected ellipticity $e$.

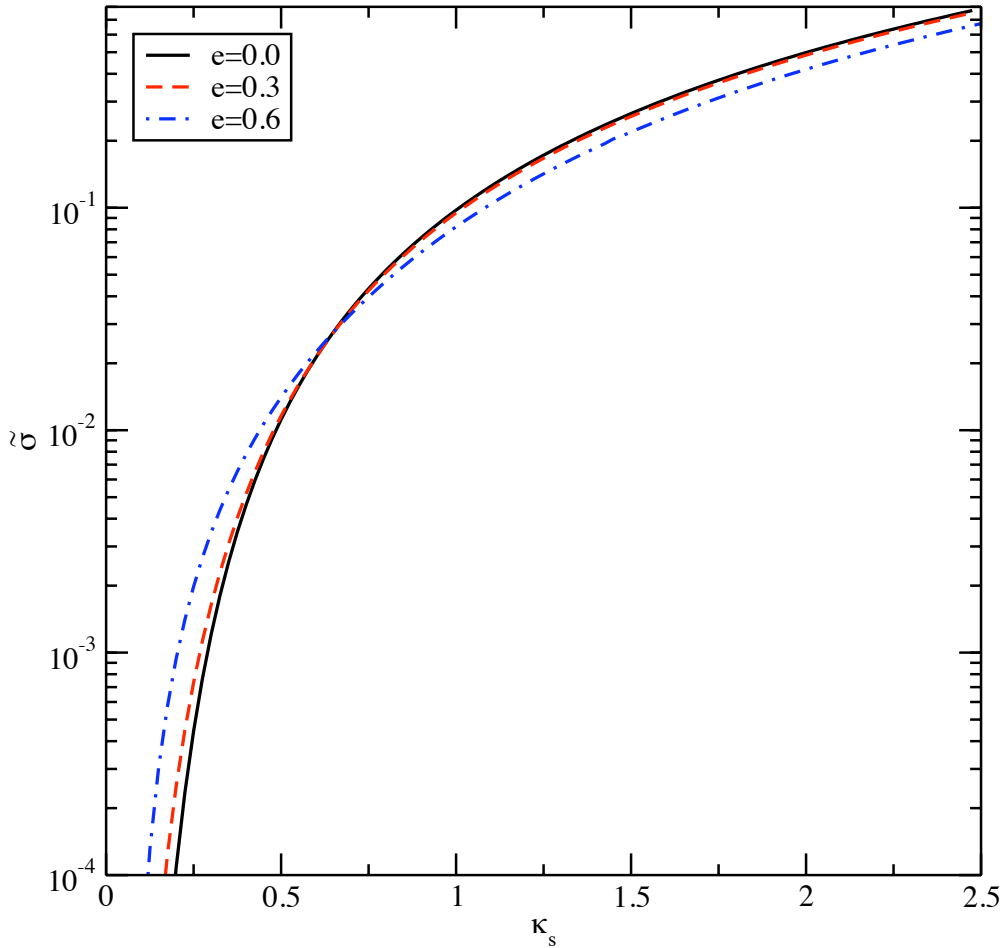

Fig. 1.— Dependence of the cross section with $\kappa_s$. A steep reduction in $\tilde{\sigma}(\kappa_s)$ is seen for low values of $\kappa_s$.

Any detailed prediction of arc abundance should take into account these effects. Nevertheless, it has been show that including finite size and source ellipticity does not change the cross section by large factors (Hattori, Watanabe, & Yamashita 1997; Bartelmann et al. 1998; Oguri 2002; Fedeli et al. 2006) and their impact is considerably smaller than lens ellipticity, for example (Oguri 2002). Li et al. (2005) made a comparison of the cross section of lensing clusters with randomly oriented elliptical sources of fixed area. They found that the cross section derived from $|\mu_t/\mu_r|$ agrees within $\sim 30\%$ with the ray-tracing simulations for $R_{\rm th} = 10$ for an average of many simulated clusters and for a generalized NFW profile. As will be shown, the impact of the magnification on arc statistics is much stronger than the above mentioned effects. Therefore, for simplicity, we will ignore the effects of finite size throughout this work and use Eq. (10) to compute the arc cross section.

In figure 1 we show the dimensionless cross section $\tilde{\sigma}$ as a function of $\kappa_s$ for three ellipticities. From this figure we see that the cross section has a sharp drop for low values of $\kappa_s$ and is, in general, not very sensitive to the ellipticity, which is in agreement with Rozo, Nagai, Keeton, & Kravtsov (2006) for distortion arcs. However, the sensitivity to $e$ increases for lower $\kappa_s$. Therefore, the ellipticity will play a more important role for clusters of lower masses and those at lower redshifts (see the results of Sect. 6).

## 5. SCALING OF THE CROSS SECTION WITH MAGNIFICATION

As discussed in Sect. 3, the magnification may have an important role in determining the observed abundance of arcs by enhancing their signal-to-noise ratio in the images. To include the effect of the magnification in $f_{arcs}$ following the method of section 3 we need to determine the cross section as a function of $\mu$. For this sake, we obtain the cross section as a function of a magnification threshold $\mu_{\rm th}$, $\tilde{\sigma}(\kappa_s, e; \mu_{\rm th})$, which is computed using the method described in section 4 above (Eq. 10) by adding the constraint that $\mu > \mu_{\rm th}$ in the integration region (besides requiring that $R > R_{\rm th}$). In figure 2 we show $\tilde{\sigma}(\kappa_s, e; \mu_{\rm th})$, for $e = 0.2$, and $\kappa_s = 0.5, 0.75, 1$ (full lines). We see that for low values of $\mu_{\rm th}$ the cross section is insensitive to this parameter. This is easy to understand, since all sources which lead to images with $R \geq R_{\mathrm{th}}$ will have a magnification above some minimum value $\mu_{\min}$. For $\mu_{\mathrm{th}} > \mu_{\min}$ the cross section begins to decrease (since not all regions that lead to $R \geq R_{\mathrm{th}}$ will have $\mu \geq \mu_{\mathrm{th}}$), exhibiting an approximately quadratic decay with $\mu_{\mathrm{th}}$ for large magnifications. This is exactly what would be expected if the cross section is dominated by arcs whose sources are close to fold caustics. Indeed, it is well known that the magnification cross section (i.e., with no constraint on $R$) scales as $\sigma \propto \mu^{-2}$ for high $\mu$ (see, e.g., Chang & Refsdal 1979; Blandford & Narayan 1986; Blandford & Kochanek 1987). In fact, for a source approaching a tangential (radial) fold, the tangential (radial) magnification $\mu_t$ ($\mu_r$) is inversely proportional to the square root of the distance between the sources and the caustic ($\beta - \beta_c$)

$$\mu_{t(r)} \propto \frac{1}{\sqrt{\beta - \beta_c}} \tag{15}$$

(see, e.g., Mollerach & Roulet 2002), which leads to the quadratic scaling for the cross section.[7]

A simple approximation for the dependence of the cross cross section on $\mu_{\mathrm{th}}$ is given by

$$\tilde{\sigma}(\kappa_s, e; \mu_{\mathrm{th}}) \simeq \tilde{\sigma}(\kappa_s, e) \times \begin{cases} 1 & \text{if } \mu_{\mathrm{th}} \leq \mu_{\mathrm{cr}}, \\ \left(\frac{\mu_{\mathrm{cr}}}{\mu_{\mathrm{th}}}\right)^2 & \text{if } \mu_{\mathrm{th}} > \mu_{\mathrm{cr}}, \end{cases} \tag{16}$$

where $\tilde{\sigma}(\kappa_s, e)$ is the cross section computed with no restriction on $\mu$ and $\mu_{\mathrm{cr}}(\kappa_s, e)$ is a critical value above which the scaling is quadratic. Notice that, by construction, the expression above is exact for $\mu < \mu_{\min}$.

The cross section for images with magnification above $\mu$ can be written as

$$\tilde{\sigma}(\kappa_s, e; \mu) = \int_\mu^\infty \frac{\mathrm{d}\tilde{\sigma}(\kappa_s, e; \mu)}{\mathrm{d}\mu} \mathrm{d}\mu \tag{17}$$

where $\mathrm{d}\tilde{\sigma}(\kappa_s, e; \mu)$ is the cross section for images with magnification between $\mu$ and $\mu + d\mu$. Using approximation (16) we have

$$\frac{\mathrm{d}\tilde{\sigma}(\kappa_s, e; \mu)}{\mathrm{d}\mu} = 2\,\tilde{\sigma}(\kappa_s, e)\frac{\mu_{\mathrm{cr}}(\kappa_s, e)^2}{\mu^3}, \tag{18}$$

if $\mu > \mu_{\mathrm{cr}}$, and zero otherwise.

Expression (16) provides an excellent approximation for $\tilde{\sigma}(\kappa_s, e; \mu)$ if $\mu_{\mathrm{cr}}$ is determined as follows. First we note that the mean magnification for images with $R > R_{\mathrm{th}}$ is given by

$$\bar{\mu} = \frac{\int_{\Omega_L(R > R_{\mathrm{th}})} \mu \, \mathrm{d}^2\theta}{\int_{\Omega_L(R > R_{\mathrm{th}})} \mathrm{d}^2\theta} = \frac{\int_0^\infty \mu \, \mathrm{d}\tilde{\sigma}(\kappa_s, e; \mu)}{\tilde{\sigma}(\kappa_s, e)}. \tag{19}$$

Then, using equation (18) we obtain

$$\bar{\mu}(\kappa_s, e) = 2\,\mu_{\mathrm{cr}}(\kappa_s, e). \tag{20}$$

The mean magnification $\bar{\mu}(\kappa_s, e)$ is computed using the definition (19). In figure 3 we show $\bar{\mu}$ as a function of $\kappa_s$ for several values of the ellipticity. The mean magnification has a steep increase for lower values of $\kappa_s$ and is very sensitive to the ellipticity. As for the dimensionless cross section, the dependence with ellipticity is inverted close at $\kappa_s \simeq 0.75$.

In figure 2 we plot the scaling of $\tilde{\sigma}$ with $\mu$, computed numerically for each magnification threshold $\mu_{\mathrm{th}}$, together with the approximation defined by equations (16) and (20). We see that they provide a good approximation in the whole $\mu$ range displayed.[8] Therefore, to obtain the cross section for a given $R_{\mathrm{th}}$ and for any specified magnification threshold we only need two functions: $\tilde{\sigma}(\kappa_s, e)$ and $\bar{\mu}(\kappa_s, e)$. In the remaining of this work we shall use this approximation to account for the behavior of the cross section as a function of the magnification threshold.

## 6. THE EFFECT OF MAGNIFICATION ON ARC STATISTICS

Now we are ready to compute the expected number of arcs per cluster including the effect of magnification following Eq. (7). The first ingredient is the dimension-full magnification-dependent cross section, which is obtained by combining Eqs. (14), (18) and (20):

$$\frac{\mathrm{d}\sigma(M, c, e, z_L; z_S; R_{\mathrm{th}}, \mu)}{\mathrm{d}\mu} = \tilde{\sigma}(\kappa_s(M, c, z_L; z_S), e; R_{\mathrm{th}}) \left[\frac{r_s(M, c, z_L)}{D_L(z_L)}\right]^2 \frac{\bar{\mu}(\kappa_s, e; R_{\mathrm{th}})^2}{\mu^3}. \tag{21}$$

The explicit dependence of $\kappa_s$ and $r_s$ in terms of the cluster mass and concentration parameter, lens and source redshifts, and cosmological model is obtained as follows. First $r_s$ and $\rho_s$ are written in terms of the concentration parameter $c$ and mass[9] $M_{200}$ as

$$r_s = \frac{r_{200}}{c} = \frac{1}{c}\left(\frac{3M_{200}}{4\pi\,200\,\rho_{\mathrm{crit}}(z)}\right)^{1/3}, \tag{22}$$

$$\rho_s = \frac{200}{3} g(c)\, c\, \rho_{\mathrm{crit}}(z), \tag{23}$$

---

[7] The contribution of cusps to the cross section scales as $\sigma \propto \mu^{-5/2}$ (Schneider & Weiss 1992; Mao 1992; Zakharov 1995). Thus, if a significant part of the cross section is contributed by cusps, this scaling has to be added to the quadratic one from folds (Schneider, Ehlers, & Falco 1992).

[8] The scaling however begins to deviate from $\mu^{-2}$ for lower $\kappa_s$ and higher ellipticities.

[9] We define $M_{200}$ as the mass enclosed in a region whose mean density contrast is 200 times the critical mass density of the Universe $\rho_{\mathrm{crit}}(z) = 3H^2(z)/(8\pi G)$ and the Hubble parameter is expressed as $H(z) = H_0\, E(z)$, with $H_0 = 100 h\, Km\, s^{-1}\, Mpc^{-1}$.

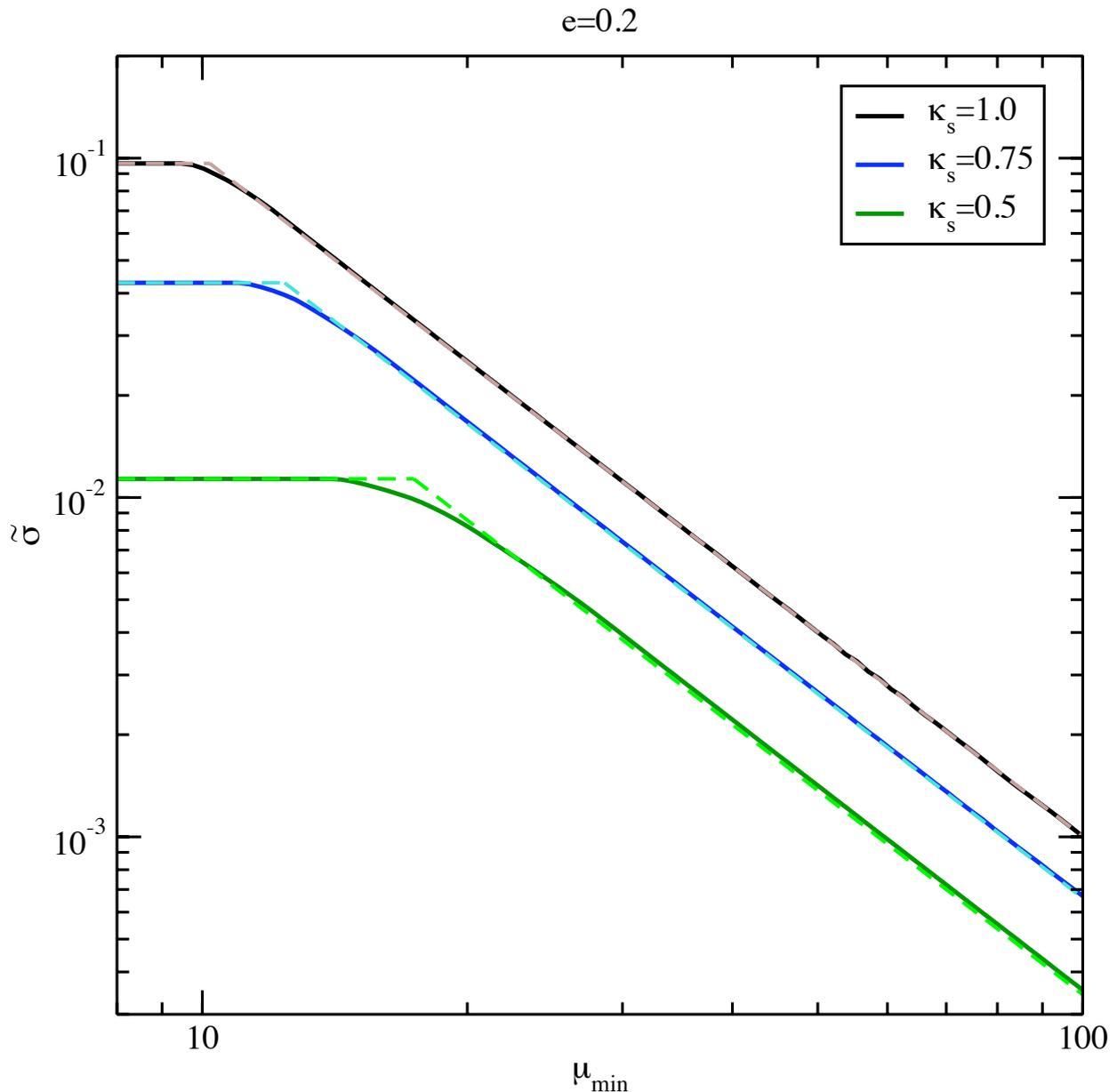

Fig. 2.— Scaling of the cross section $\tilde{\sigma}(\kappa_s, e; \mu_{\rm th})$, with respect to the magnification threshold $\mu_{\rm th}$, for the elliptic NFW model with $e = 0.2$. The full lines show the numerical computation using the conditions $R > R_{\rm th}$ and $\mu > \mu_{\rm th}$ in equation (10). The dashed lines show the results of the approximation defined by equations (16) and (20). The vertical lines indicate the values of $\mu_{\rm min}$ and $\mu_{\rm crit}$ for $\kappa_s = 0.5$.

where $g(c) = c^2 / (\ln(1 + c) - c / (1 + c))$.

Obviously the relation (22) among the mass and radius holds for a spherical cluster. Our elliptical clusters are obtained by "deforming" the mass distribution in the direction perpendicular to the line of sight according to equation (13). In other words, a region of constant density would define a shell $\xi^2 + z^2 = const.$ Since, by construction, the area of a $\xi = const.$ curve is independent of $e$, the mass is independent of the ellipticity and equation (22) will hold even in the elliptical case. It is worth stressing that, by this procedure, the elliptical clusters are oblate spheroids with the smaller axis perpendicular to the line of sight. We are thus neglecting the triaxiality of these objects. However, this situation is somewhat intermediate to having a triaxial cluster aligned along the major or minor axes. For a full treatment of triaxiality, see e.g., Oguri, Lee, & Suto (2003). For the purposes of this paper it is sufficient to proceed with the approximation described above.

Then, using Eqs. (12), (22), and (23), and inserting the relevant units, the projected NFW profile parameters are

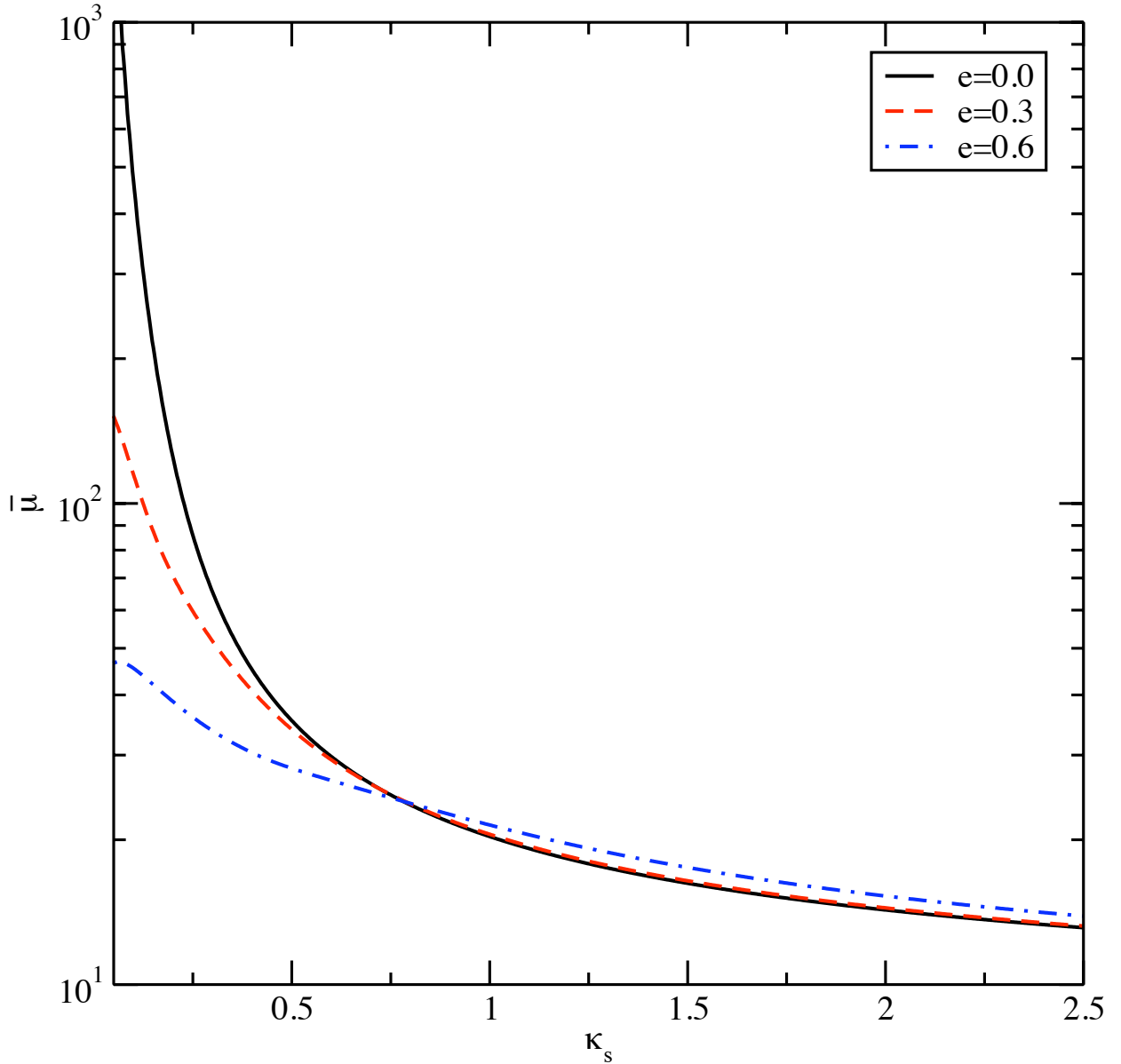

FIG. 3.— Average magnification within the region with $|\mu_t/\mu_r| > R_{\mathrm{th}} = 10$ as a function of $\kappa_s$ for several values of the ellipticity.

finally given by

$$\kappa_s = 7.36 \times 10^{-4} g(c) \frac{(200 E^2(z_L))^{2/3}}{1 + z_L} \left( \frac{M_{200} \, h}{10^{14} \, M_\odot} \right)^{1/3} \frac{I_{OL} \, I_{LS}}{I_{OS}}, \tag{24}$$

$$x_s = \frac{r_s}{D_{OL}} = \frac{1.47 \times 10^{-3}}{I_{OL} \, c} \frac{1 + z_L}{(200 E^2(z_L))^{1/3}} \left( \frac{M_{200} \, h}{10^{14} \, M_\odot} \right)^{1/3}, \tag{25}$$

where $I_{LS}$ is the transverse commoving distance from lens to source in units of the Hubble radius (and equivalently for $I_{OL}$ and $I_{OS}$): $I_{LS} = S_k \left( \sqrt{|\Omega_k|} \, \chi_{LS} \right) / \left( \sqrt{|\Omega_k|} \right)$, with $\chi_{LS} = \int_{z_L}^{z_S} \frac{dz}{E(z)}$ and $S_k(\chi) = \sin\chi, \chi, \sinh\chi$, for $\Omega_k < 0, = 0, > 0$, respectively. Here we adopt a flat $\Lambda$CDM model, such that $E(z) = \sqrt{\Omega_m(1+z)^3 + \Omega_\Lambda}$ and $\Omega_k = 1 - \Omega_m - \Omega_\Lambda = 0$, with parameters $\Omega_m = 0.3$, $\Omega_\Lambda = 0.7$. Notice that if the mass is expressed in units of $h^{-1}$, the results will be independent on the Hubble parameter.

For a given background cosmological model (and fixed $z_L$ and $z_S$), $\kappa_s$ and $r_s$ are functions of $M_{200}$ and $c$ only. Although $c$ is in principle independent of $M_{200}$, these two quantities are strongly correlated, as seen in simulations (see, e.g. Klypin, et al. 1999) and in observational data (Comerford & Natarajan 2007). Here we use the expression for $c(M_{200})$ from Neto et al. (2007), with the redshift dependence given in Macciò et al. (2008). Although the distribution of halo concentrations for a given mass and redshift follows a log-normal distribution, we shall use only the mean relation in this work.

The second ingredient to compute the arc fraction is the source distribution as a function of redshift and limiting magnitude. The redshift dependence is modelled as in Baugh & Efstathiou (1993) and we take the total surface density of sources and the scaling with $m_{\lim}$ from the COMBO-17 survey (Taylor et al. 2007), such that

$$n(z_S, m_{\lim}) \simeq \frac{30}{\text{arcmin}^2} \left[z_m(m_{\lim})\right]^{3.4} \times \frac{3 z_S^2}{2 \left[z_*(m_{\lim})\right]^3} \exp\left[-\left(\frac{z_S}{z_*(m_{\lim})}\right)^{3/2}\right],  \tag{26}$$

where $z_* = z_m/1.412$ and $z_m$ is the median redshift, which depends on the limiting magnitude $m_{\lim}$ in the relevant band. For an $R$-band survey $z_m \simeq 0.23(m_{\lim} - 20.6)$ (Brown et al. 2003).

Although we do not explicitly take into account the effects of the PSF in this work, we point out that the distribution above was obtained from ground-based observations. Therefore, our arc fraction computation will include, to some extent, the effect of the seeing. Ideally one would determine the source distribution from deep space based observations and include the effects of the PSF through the formalism of Sect. 2. This is left for future work.

With the relations above, we may finally compute the expected number of arcs per cluster, which is given by a convolution of the cross section with the source surface density, integrated over all sources behind the cluster and all the magnifications (Eq. 7):

$$f_{arcs}(M, e, z_L) = \int_{z_L}^{\infty} \int_{\bar{\mu}/2}^{\infty} \frac{\mathrm{d}\sigma(M, e, z_L, z_S; \mu, R_{\text{th}} = 10)}{\mathrm{d}\mu} \times n\left(z_S, m_{\lim}^{\text{eff}}(\mu, m_{\lim})\right) \mathrm{d}\mu \, \mathrm{d}z_S. \tag{27}$$

We also test an approximation by which the whole area defined by the condition $R > R_{\text{th}}$ has the mean magnification $\bar{\mu}$ (Eq. 19), such that $\tilde{\sigma}(\kappa_s, e; \mu_{\text{th}}) = \tilde{\sigma}(\kappa_s, e)\theta(\mu - \bar{\mu})$. In this case, Eq. (27) simplifies to

$$f_{arcs}(M, e, z_L) = \int_{z_L}^{\infty} \sigma(M, e, z_L, z_S; R_{\text{th}} = 10) \times n\left(z_S, m_{\lim}^{\text{eff}}(\bar{\mu}, m_{\lim})\right) \mathrm{d}z_S, \tag{28}$$

where $\sigma$ is the dimension-full cross section without imposing any cut in magnification. In other words, all sources are assumed to be magnified by the mean magnification $\bar{\mu}$. As above, the dependence on $z_L$, $z_S$, and $M$ enters on both $\sigma$ and $\bar{\mu}(\kappa_s, e; R_{\text{th}})$ through $\kappa_s$.

For the sake of comparison, we also compute the arc fraction neglecting the effect of magnification, i.e., using $m_{\lim}^{\text{eff}} = m_{\lim}$ in the equation above, which is the more standard computation. In figure 4 we show the results of these 3 computations for the arc fraction $f_{arcs}$ as a function of lens redshift for a cluster with $M = 5 \times 10^{14} h^{-1} M_{\odot}$ and $e = 0.5$ and a survey with limiting magnitude $R_{\lim} = 24$. We see that the magnification has a dramatic impact on arc abundance, enhancing the fraction of arcs by more than an order of magnitude. Furthermore, the inclusion of the effect of magnification shifts the peak of the arc incidence to higher redshifts and modifies the slope of the decay at large $z_L$, showing a softer decrease with lens redshift.

We also notice that the results including the dependence of the cross section with $\mu$ as in Eqs. (18) and (27) (full line in Fig. 4) or considering only the mean magnification, as in Eq. (28) (dashed line), are very similar. Indeed, the relative difference is less than 10% for a broad range of parameters. Therefore, from now on we will use the *constant magnification approximation* given by (28), since it makes calculations much faster, by avoiding the integration in $\mu$.

In Fig. 5 we show the impact of the depth of the survey on $f_{arcs}(z)$ for the same lens parameters as in the previous figure. The bottom line corresponds to $R_{\lim} = 22$, typical of the Sloan Digital Sky Survey,[10] while the upper line corresponds to $R_{\lim} = 28$, close to what would be achieved from the final Large Synoptic Survey Telescope[11] (LSST) coadded data (Ivezić et al. 2008). As expected, the change in limiting magnitude has a large impact on arc incidence, which could explain the differences in arc abundance found in the various surveys. Also, by comparing Figs. 4 and 5, it is clear that the effect of magnification is qualitatively akin to changing the limiting magnitude of the survey.

In Fig. 6 we show the result of the integration of Eq. (28) as a function of $z_L$ for several values of $M$ and $e$. The qualitative behaviour is similar in all cases, with the expected number of arcs increasing as a function of cluster redshift at low redshifts and decreasing at higher redshifts. These results can be simply understood in terms of the scaling of the cross section with the characteristic convergence $\kappa_s$ (Fig. 1), the scaling of $\kappa_s$ with mass and redshift (Eq. 24), and the source redshift distribution (Eq. 26). First, for a given mass and ellipticity, the maximum value of $\kappa_s$ increases with $z_L$. In other words, the lensing efficiency has the potential to grow with $z_L$, as long as there are far enough sources to be lensed. Therefore, for low lens redshifts, $f_{arcs}$ will always increase with $z_L$. On the other hand, for high $z_L$ the exponential decay of sources will end up causing a fall down in $f_{arcs}$. The arc fraction is more sensitive to $z_L$ for lower masses, ellipticities, and lens redshifts. This occurs because the cross section is much more sensitive to $\kappa_s$ for smaller values of this parameter and smaller ellipticites.

From these figures we see that the arc fraction is more sensitive to the ellipticity than to the mass, with roughly a linear dependence on mass at the peak and exponential with ellipticity. The transition redshift (i.e. the redshift at which $f_{arcs}$ starts to decrease) is weakly dependent on $M$ and $e$.

The dependence of arc abundance on the source distribution and its dependence on the limiting magnitude highlight the importance of the magnification effects for $f_{arcs}$. For example, for $\kappa_s \simeq 1$, we have $\mu_{\min} \simeq 20$ (see Fig. 3), which implies that the effective limiting magnitude is increased by at least 1.25 (see Eq. 4), changing both the normalization and shape of $f_{arcs}(z_L)$.

[10] http://www.sdss.org/
[11] http://www.lsst.org

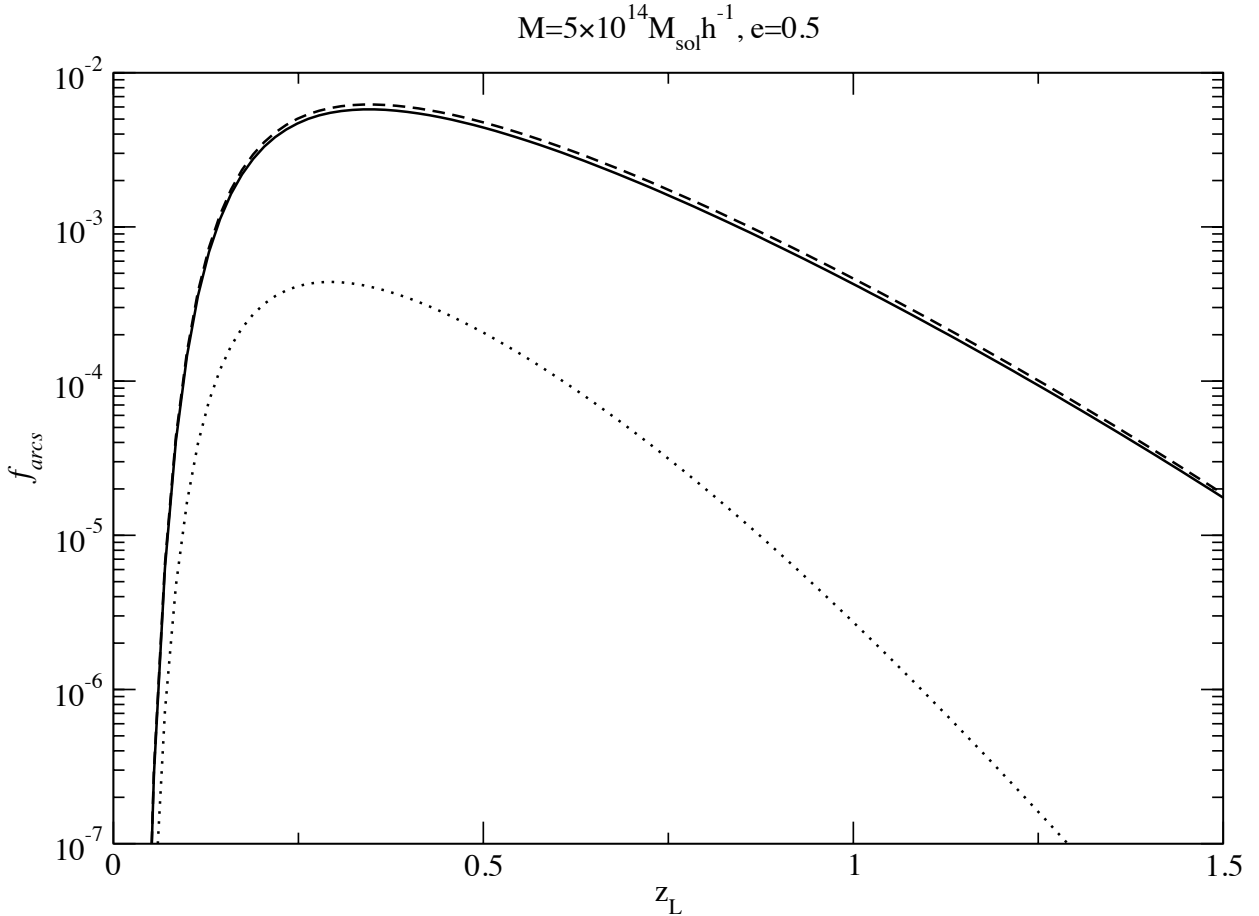

Fig. 4.— The expected fraction of arcs as a function of lens redshift computed in three different ways: ignoring the magnification (dotted line), taking into account the mean magnification (dashed line), and considering the quadratic scaling of the cross section with magnification (full line). The results are shown for a NFW cluster with $M = 5 \times 10^{14} h^{-1} M_\odot$ and $e = 0.5$, assuming $R_{\mathrm{lim}} = 24$.

## 7. CONCLUDING REMARKS

In this paper we have introduced a method to include the effect of noise and other observational effects in semi-analytic computations of gravitational arc statistics. We argue that the expected fraction of arcs per cluster $f_{arcs}$ is a well suited indicator to study arc statistics. On the theoretical side this quantity is less dependent on cosmological parameters not directly related to lensing (such as $\sigma_8$) than the total arc number. On the observational side $f_{arcs}$ is less affected by selection effects. The procedure to compute $f_{arcs}$ involves the arc cross section $\sigma$, the source distribution, and a combination of observational effects with the selection function (see Eq. 1). In particular, to include the effect of the increased $S/N$ of the arcs due to the magnification on their detectability, we are lead to consider the magnification dependent cross section. This cross section is obtained by adding an additional constraint on the image properties. Besides requiring that $L/W > R_{\mathrm{th}}$, we also impose that the sources must be magnified by a minimum factor $\mu_{\mathrm{th}}$.

The cross section was computed through a semi-analytic method based on the local eigenvalue ratio of the magnification tensor. This method has been used in previous studies (see, e.g., Hattori, Watanabe, & Yamashita 1997; Fedeli et al. 2006), and was validated using simulations of extended sources. For distortion arcs, the local eigenvalue ratio gives a good approximation for the cross section and is much faster than ray tracing finite sources. Using this method we investigated the scaling of $\sigma$ with respect to the magnification threshold $\mu_{\mathrm{th}}$. We found that $\sigma$ scales approximately as $\mu_{\mathrm{th}}^{-2}$ for high magnifications — which is expected from the properties of the lens mapping close to folds — and obtained a good approximation for all values of $\mu_{\mathrm{th}}$ using the mean magnification $\bar{\mu}$. This approximation is expected to be valid for generic lens models (as long as the scaling is close to the one expected by fold arcs).

Given a cluster mass model, the cosmological dependence of the cross section is encoded on the projected characteristic density and characteristic scale. In the case of the elliptical NFW profile, the problem of computing $\sigma(M, e, z_L; z_S)$ is simplified to the one of obtaining $\tilde{\sigma}(\kappa_s, e)$. Further, if one wishes to approximate the cross section for arbitrary $\mu_{\mathrm{th}}$ it suffices to obtain $\tilde{\sigma}(\kappa_s, e)$ and $\bar{\mu}(\kappa_s, e)$. Obviously, the scaling of $\sigma$ with projected mass and (arbitrary) length scale is a property of the lensing geometry. This feature could be used to scale the cross sections of simulated clusters for different lens and source redshifts if the lens parameters are scaled accordingly.

To obtain the arc fraction we assume that the objects are identified and selected in terms of their $S/N$ in the images. The arc selection function is given by a constant threshold in $L/W$ and $S/N$. The enhanced detectability due to the magnification acts effectively as an increased depth of the image in the magnified region, which translates into a (magnification dependent) modification of the source distribution. We account for this effect by effectively changing the limiting magnitude used to determine the source distribution as in equation (4). We take the source surface density as a function of redshift and its scaling with $m_{\lim}$ from the COMBO-17 survey.

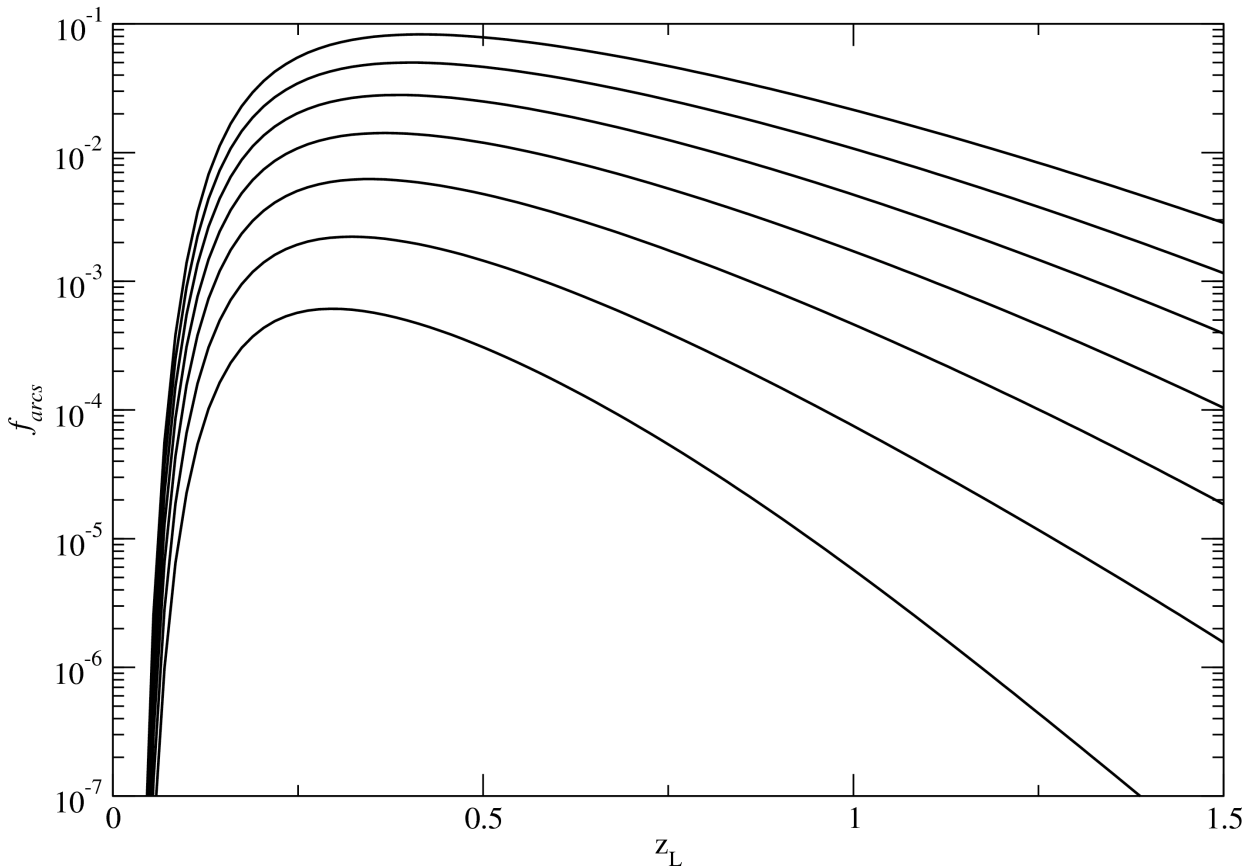

FIG. 5.— The expected fraction of arcs as a function of lens redshift for different limiting magnitudes for a NFW cluster with $M = 5 \times 10^{14} h^{-1} M_\odot$ and $e = 0.5$. The bottom line shows the results for $R_{\lim} = 22$, increasing in steps of unity, to reach $R_{\lim} = 28$ for the upper curve.

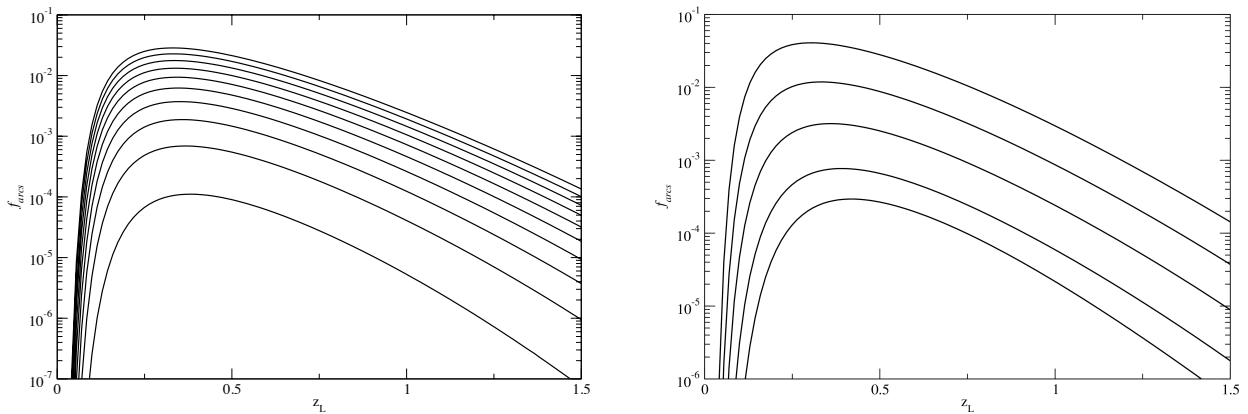

FIG. 6.— Expected fraction of arcs as a function of lens redshift for different masses and ellipticities for a NFW cluster in a survey with $R_{\lim} = 24$. Left: fixed ellipticity ($e = 0.5$) and varying the mass in steps of $10^{14} h^{-1} M_\odot$, from $M = 1 \times 10^{14} h^{-1} M_\odot$ (lower curve) up to $M = 10^{15} M_\odot$ (upper curve). Right: fixed mass ($M = 5 \times 10^{14} h^{-1} M_\odot$) and varying ellipticity, from $e = 0$ (bottom line), increasing in steps of 0.2 up to $e = 0.8$ for the upper curve.

The resulting arc fraction is shown as a function of $z_L$ in figure 4. We see that the inclusion of the magnification has a very strong impact on the arc incidence, both enhancing the arc fraction by over an order of magnitude, as well as changing the shape of its redshift distribution. In particular, the peak in the arc fraction shifts towards higher $z_L$ and the tail of $f_{arcs}$ has a less steep decrease with the cluster redshift. All these features are qualitatively in the right direction for improving the agreement between the predictions and observations of arc abundance, though the modelling used in this work is still oversimplified to allow for a direct comparison. Also, the detailed form and magnitude of the arc fraction is sensitive to the redshift variation of the mass-concentration relation, the mass definition, etc. Besides, it must be pointed out that the impact of the magnification may be a bit overemphasised in these results, given our assumption on the direct dependence of detection on the arc $S/N$. Other quantities may also play an important role, such as the surface brightness, which is not directly affected by the magnification. In any case, we emphasise the importance of including the effect of magnification, which is often ignored in semi-analytic predictions of arc abundance and is not accounted for in simulations that use constant surface brightness sources.

The *magnification bias* considered in this work is different from the standard magnification bias in the statistics of multiple imaging of point sources (see, e.g., Turner 1980; Turner et al. 1984), in that the effect scales with the square root of magnification (i.e. $S/N$ enhancement) and not the magnification itself. The magnification has been considered in previous arc statistics studies (see, e.g., Oguri, Lee, & Suto 2003), but through the enhancement of the total flux, which should not be the dominant factor for arc detection.

As far as we know, this is the first work to take into account the effect of the magnification through the $S/N$ on arc statistics and to consider the dependence of the cross section on the magnification for computing the arc abundance. We compute the arc fraction using two methods: by using the magnification dependent cross section (Eq. 27) and by using the cross section with no cut in $\mu_{\rm th}$ but associating to all arcs the mean magnification $\bar{\mu}$ (Eq. 28). Both methods turn out to be in very good agreement. Therefore, it is sufficient to consider the magnification independent cross section and carry out the integration only on the source redshift, as in the more standard approach, as long as we consider the effect of the mean magnification (which depends on the lens parameters and source redshift) on the source distribution. The inclusion of magnification on arc statistics through this procedure requires only the additional computation of $\bar{\mu}$ in addition to the cross section.

Several aspects of this work could be improved to yield more realistic predictions for the arc cross section and the expected number of arcs per cluster. Regarding the lens model, two key features that should be considered are the triaxial mass distribution (see, e.g., Oguri, Lee, & Suto 2003; Hennawi, Dalal, Bode, & Ostriker 2007) and the mass-concentration relation $c(M)$, including its scatter. More sophisticated mass distributions could also be dealt with, including different radial profiles, external perturbations, asymmetries, and substructures. Regarding the sources, the effects of finite sizes and ellipticities could be included with the methods discussed in section 4. Finally, given the observed morphological properties of the sources (e.g. their size distribution), the effects of the seeing should be taken into account, as outlined in section 2. This is left for future work. In any case, we argue that the enhancement of the $S/N$ of the arcs due to the magnification must be included in more sophisticated semi-analytic modelling of arc abundance. Furthermore, the inclusion of this effect could help to bring a better agreement between observations and theoretical modelling of arc abundance.

With the upcoming wide-field surveys such as DES, which has just started its first observing season, and LSST, and future space based projects such as EUCLID[12] (Laureijs et al. 2011), the observed number of gravitational arcs is expected to reach $\sim 10^3$–$10^4$. This will become a promising new era for doing precision statistical studies of the cluster inner matter distribution and its evolution through strong lensing. For such a program to be successful, end-to-end simulations to generate fake images are starting to be extensively used in combination with arcfinding codes, allowing for a direct comparison of theoretical predictions with observational data, taking into account the physical processes involved and the observational effects. As in many areas of cosmology, the interplay between detailed simulations and simple semi-analytic models is a useful and necessary step. The semi-analytic approach allows for a qualitative understanding of the problem and provides a framework to interpret the outcome of simulations. Furthermore, the semi-analytic modelling provides a way to make fast computations allowing a large parameter space to be probed for performing parameter estimations and forecasts.

### ACKNOWLEDGMENTS

The authors are grateful to Silvia Mollerach and Esteban Roulet for providing fundamental contributions to this work in its earlier stages. We thank Eduardo Rozo for numerous remarks that lead to a significant improvement of this work. The authors acknowledge Chuck Keeton for making his `gravlens` code publicly available, which was used to test our numerical codes. MM wishes to thank Eduardo Cypriano, Laerte Sodré, and Joan-Marc Miralles for useful comments and suggestions. JE and MM acknowledge the hospitality of Instituto Balseiro, where this work was initiated. MM acknowledges the hospitality of the Particle Astrophysics Center at Fermilab were part of this work was done. GBC is partially funded by CNPq and CAPES. MM is partially supported by CNPq (grant 309804/2012-4) and FAPERJ (grant E-26/110.516/2012).

[12] http://www.euclid-ec.org/

APPENDIX: LENS POTENTIAL DERIVATIVES FOR ARBITRARY ELLIPTICITY PARAMETERIZATION

An elliptical convergence can be obtained from a circular symmetric one by replacing the radial coordinate by

$$\xi^2 = \frac{x_1^2}{a^2} + \frac{x_2^2}{b^2}, \tag{29}$$

where $a$ and $b$ are parameters related to the convergence ellipticity. Different parameterizations can be adopted, for instance, $a = 1$, $b = (1 - e)$ (see, e.g., Keeton 2001b); $a = 1/(1 - e)$, $b = 1 - e$ (see, e.g., Lima et al. 2010); $a = 1 - e$, $b = 1 + e$ (Blandford & Kochanek 1987; Golse & Kneib 2002; Dúmet-Montoya et al. 2012), or $a = (1 - e)^{-1/2}$ and $b = (1 - e)^{1/2}$, which is the convention we adopt in this work (since it keeps constant the area within the ellipses, and therefore the enclosed mass, when varying $e$).

Schramm (1990) derived the angle deflection for elliptical convergences in terms of integrals as follows:

$$\hat{\alpha}_1(x_1, x_2) = -\frac{8\pi G}{c^2}ab \int_0^{m(x_1,x_2)} \frac{p'^2}{a'b'} \frac{x_1}{a'^2}\Sigma(m)\,dm, \tag{30}$$

$$\hat{\alpha}_2(x_1, x_2) = -\frac{8\pi G}{c^2}ab \int_0^{m(x_1,x_2)} \frac{p'^2}{a'b'} \frac{x_2}{b'^2}\Sigma(m)\,dm, \tag{31}$$

where $p'$, $a'$, $b'$ and $m$ are defined in Schramm (1990), and $\Sigma(r)$ is the projected mass density of the axially symmetric model. Following Keeton (2001b), it is useful to make the following variable substitution to simplify these integrals:

$$m^2 = u\left(\frac{x_1^2}{1 - (1 - a^2)u} + \frac{x_2^2}{1 - (1 - b^2)u}\right). \tag{32}$$

Then, computing the partial derivatives of the deflection angle, is possible to write:

$$\Psi_1(x_1, x_2) = abx_1 J_0, \tag{33}$$
$$\Psi_2(x_1, x_2) = abx_2 J_1, \tag{34}$$
$$\Psi_{11}(x_1, x_2) = 2abx_1^2 K_0 + abJ_0, \tag{35}$$
$$\Psi_{22}(x_1, x_2) = 2abx_2^2 K_2 + abJ_1, \tag{36}$$
$$\Psi_{12}(x_1, x_2) = 2abx_1 x_2 K_1, \tag{37}$$

where

$$J_n(x_1, x_2) = \int_0^1 \frac{\kappa(m(u))}{[1 - (1 - b^2)u]^{\frac{1}{2}+n}[1 - (1 - a^2)u]^{\frac{3}{2}-n}}du, \tag{38}$$

$$K_n(x_1, x_2) = \int_0^1 \frac{u\kappa'(m(u))}{[1 - (1 - b^2)u]^{\frac{1}{2}+n}[1 - (1 - a^2)u]^{\frac{5}{2}-n}}du, \tag{39}$$

and $\kappa'(m) = \frac{d\kappa(m)}{d(m^2)}$. Fixing $a = 1$, $b = (1 - e)$ yields the same expressions as in Keeton (2001b).

From equations $(33) - (37)$ it is possible to compute directly some quantities related to the lensing, for instance, the tangential and radial magnifications are obtained through by Eqs. (8) and (9).

The expressions above can be evaluated using standard numerical methods and libraries, such as Gaussian quadrature integrations (see, e.g., section 4.5 of Press et al. 1992) and the `gsl` library. We have tested our numerical results with the well established code `gravlens` by comparing the numerical results for the lens potential derivatives (Eq. 33–37), magnification, and the deformation cross section for a few lens models. Furthermore, for $\kappa_s \lesssim 0.1$ accurate analytic approximations can be found for these lensing quantities (Dúmet-Montoya et al. 2013).

The numerical codes used for this work are part of the *Strong Lensing Tools* (`SLtools`, `http://icra.cbpf.br/sltools/`) library and are available upon request to the authors.